# Nucleon-induced fission cross-sections of tantalum and separated tungsten isotopes and "compound nucleus" effect in intermediate energy region

A. N. Smirnov[1,a], O. I. Batenkov[1], V. P. Eismont[1], N. P. Filatov[1], J. Blomgren[2], H. Condé[2], A. V. Prokofiev[3] and S. G. Mashnik[4]

[1]V.G. Khlopin Radium Institute, 2-nd Murinskiy prospect 28, Saint-Petersburg 194021, Russia
[2]Department of Neutron Research, Uppsala University, Box 525, S-751 20 Uppsala, Sweden
[3]The Svedberg Laboratory, Uppsala University, Box 533, S-751 21 Uppsala, Sweden
[4]Los Alamos National Laboratory, Los Alamos, NM 87545, USA

**Abstract.** Neutron- and proton-induced fission cross-sections of separated isotopes of tungsten ($^{182}$W, $^{183}$W, $^{184}$W, and $^{186}$W) and $^{181}$Ta relative to $^{209}$Bi have been measured in the incident nucleon energy region 50 – 200 MeV using fission chambers based on thin-film breakdown counters (TFBC) using quasi-monoenergetic neutrons from the $^7$Li(p,n) reaction and at the proton beams of The Svedberg Laboratory (TSL), Uppsala University (Uppsala, Sweden). The results are compared with predictions by the CEM03.01 event generator, as well as with the recent data for nuclei in the lead-bismuth region. The effect of "compound nucleus" in the intermediate energy region is discussed, displaying in exponential dependence of nucleon-induced fission cross-sections on the parameter $Z^2/A$ of the composite system (projectile+target nucleus), and in other characteristics of the fission process for which parameter $Z^2/A$ plays a role similar to the one of the usual liquid-drop parameter $Z^2/A$ of compound nuclei.

## 1 Introduction

Fission cross-sections of W (isotope $^{184}$W) in the energy region up to the 200 MeV was included in the High Priority Request List [1] due to its importance for development of ADS and nuclear reaction models. Recently, evaluated cross-section data libraries have been created at LANL for all stable tungsten isotopes up to 150 MeV [2]. However, there are no evaluated data sets for fission cross-sections. The experimental database up to now consisted only of data for $^{nat}$W(n,f) cross sections in the energy range 50 – 200 MeV [3,4] and $^{nat}$W(p,f) one at 190 MeV [5].

Our preliminary results on fission cross sections for separated tungsten isotopes have already been published in ref. [6] together with calculation by CEM03.01 generator [7]. In this work the final experimental data are presented. In addition, they are supplemented by our recent results for $^{181}$Ta. The results are analyzed in comparison with the (n,f) and (p,f) cross-sections data for bismuth and separated isotopes of lead in the 50-200 MeV energy range.

General regularities are discussed in some characteristics of the fission process in the intermediate energy region for which the parameter $Z^2/A$ of the composite system (incident nucleon + target nucleus) plays a role similar to the one of the usual liquid-drop parameter $Z^2/A$ of compound nuclei.

## 2 Experimental results

The measurements have been carried out at the upgraded neutron beam facility [8] and at the broad proton beam facility [9] of the TSL. Similar to our previous work [10], the same fission chambers based on thin-film breakdown counters (TFBC) were used at both neutron and proton fluxes Fissile targets were made from separated isotopes of tungsten in the WO$_3$ chemical form by means of vacuum evaporation onto aluminum backings. The thicknesses of the deposited layers, about 2 mg/cm$^2$, have been defined by direct weighting and by method of Rutherford backscattering of α-particles [11]. The measurements of relative counting rates of fission events induced by neutrons were carried out for all separated tungsten isotopes simultaneously. Measurements in wide proton beam were carried out simultaneously for two or three target sets only (depending on proton energy) while one of them was used as a relative monitor of the proton flux. The total statistical uncertainties of the relative results depend on the projectile energy and amount to not more than 10% for the neutron measurements. For proton measurements, they vary from 5-10% above 90 MeV to 15-25% below 60 MeV. In the data reduction process, the corrections were introduced, connected with:

− the TFBC detection efficiency, taking into account of properties of fission fragments,

− the shape of the neutron spectrum (for the neutron part

---

[a] Presenting author, e-mail: smirnov@atom.nw.ru

of the measurements);

- the background of fission events due to the presence of heavy fissile admixtures in the aluminum backings.

The main aspects of the data processing have been described in details in refs. [3, 12]. Absolute (n,f) and (p,f) cross-sections for separated isotopes of tungsten ($^{182,183,184,186}$W) and $^{181}$Ta have been obtained by multiplication of the measured cross-section ratios to parameterized $^{209}$Bi(n,f) [3] and $^{209}$Bi(p,f) [10] cross-section, respectively. In view of their practical importance, we present also the (n,f)- and (p,f) cross-section data for $^{nat}$W obtained both by direct measurements using the $^{nat}$W samples and calculations on a basis of results for separated isotopes taking into account their contributions in the natural element. The results are presented in Tables 1 and 2 and are shown in fig. 1 (only for tungsten isotopes and Ta).

**Table 1.** Neutron-induced fission cross-sections.

| $E_n$ MeV | $^{182}$W mb | $^{183}$W mb | $^{184}$W mb | $^{186}$W mb | $^{nat}$W exp. mb | $^{nat}$W calc. mb |
|---|---|---|---|---|---|---|
| 46.6±1.0 | ≤ 0.003 | | ≤ 0.006 | ≤ 0.0016 | <0.0040 | |
| 65.1±1.7 | 0.035±0.004 | 0.027±0.003 | 0.021±0.002 | 0.009±0.001 | 0.020±0.010 | 0.020±0.002 |
| 74.1±1.4 | 0.071±0.009 | 0.065±0.008 | 0.039±0.005 | 0.023±0.007 | 0.042±0.009 | 0.046±0.007 |
| 94.6±1.2 | 0.26±0.03 | 0.19±0.02 | 0.14±0.02 | 0.071±0.008 | 0.14±0.02 | 0.16±0.02 |
| 142.5±3.1 | 1.0±0.1 | 0.64±0.09 | 0.62±0.08 | 0.33±0.05 | 0.56±0.08 | 0.64±0.09 |
| 172.4±3.0 | 1.4±0.2 | 1.16±0.13 | 1.0±0.1 | 0.48±0.06 | 1.04±0.15 | 0.96±0.11 |

**Table 2.** Proton-induced fission cross-sections.

| $E_p$ MeV | $^{182}$W mb | $^{183}$W mb | $^{184}$W mb | $^{186}$W mb | $^{181}$Ta mb | $^{nat}$W exp. mb | $^{nat}$W calc. mb |
|---|---|---|---|---|---|---|---|
| 40.4±0.2 | 0.0048±0.0007 | 0.0045±0.0007 | 0.0041±0.0006 | 0.0026±0.0004 | - | - | 0.0039±0.0006 |
| 43.5±0.9 | 0.011±0.002 | 0.0104±0.0017 | 0.0087±0.0014 | 0.0045±0.0008 | - | - | 0.008±0.001 |
| 59.2±1.0 | 0.09±0.02 | 0.07±0.01 | 0.039±0.007 | 0.024±0.005 | - | 0.050±0.006 | 0.054±0.006 |
| 64±1.0 | - | - | - | - | 0.039±0.017 | - | - |
| 93.2±1.0 | 1.24±0.17 | 0.88±0.12 | 0.62±0.09 | 0.34±0.04 | 0.36±0.07 | 0.71±0.10 | 0.74±0.10 |
| 130.5±1.0 | 2.93±0.35 | 2.2±0.3 | 1.8±0.2 | 0.98±0.12 | 0.94±0.18 | 1.98±0.23 | 1.92±0.23 |
| 166±1.0 | - | - | - | - | 1.53±0.23 | - | - |
| 169.6±1.0 | 4.85±0.58 | 3.9±0.5 | 3.2±0.4 | 2.0±0.2 | 1.91±0.28 | 3.5±0.4 | 3.4±0.4 |

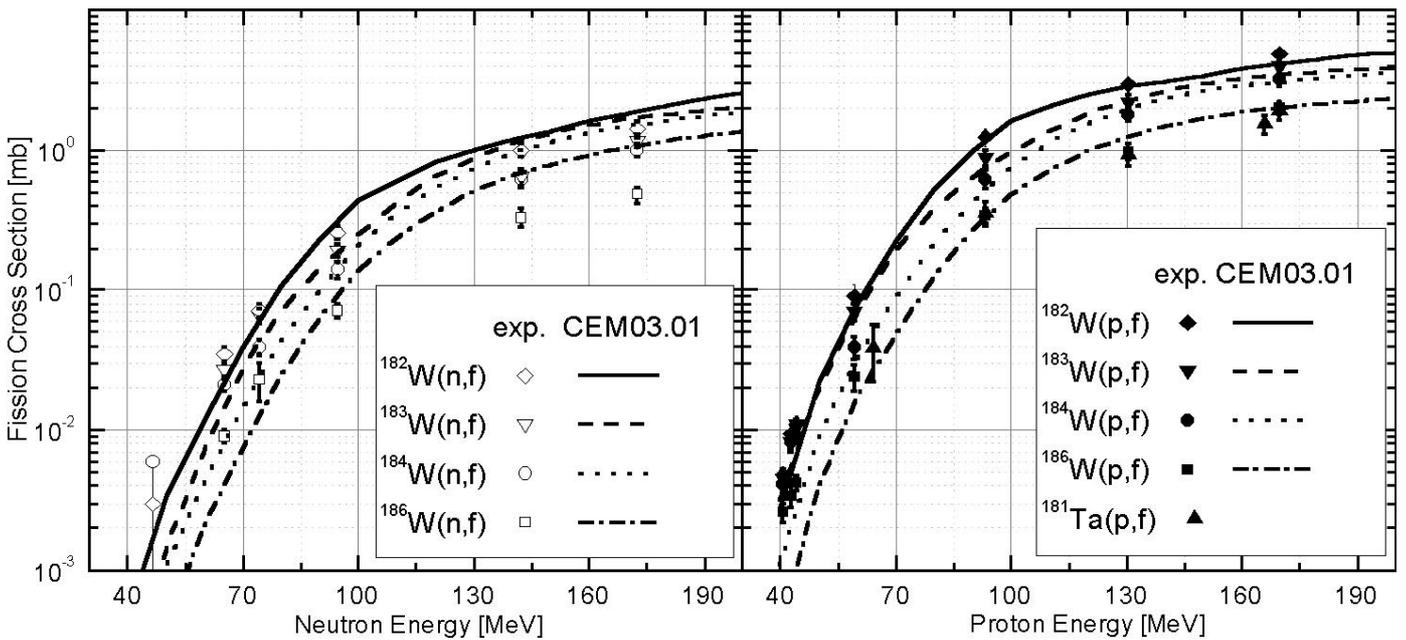

Fig. 1. The (n,f) and (p,f) cross-sections of $^{182}$W, $^{183}$W, $^{184}$W, $^{186}$W and $^{181}$Ta(p,f) versus incident nucleon energy. Symbols are experimental results of the present work, and curves are results of calculations by the CEM03.01 event generator.

In Fig. 1 the experimental data are compared with the results of updated calculations for tungsten isotopes by the CEM03.01 event generator [7]. It is seen that the calculations agree satisfactory with the (p,f) data, while problems are seen in description of the (n,f) data. At present this is an open questions of CEM03.01 still to be solved.

## 3 Isotopic dependence of the cross sections

A low contribution of the shell correction (~1 MeV) to the fission barrier (~30 MeV) and a large deformation (β≈0.25) for the tungsten isotopes distinguish them significantly from the lead isotopes studied earlier. The latter are of spherical shape and have the largest microscopic contribution to the barrier of all nuclei (~15 MeV). Thus, the tungsten isotopes can be considered as representatives of the typical liquid-drop fission.

It is seen from the fig. 1 that the values of the (n,f) and (p,f) cross sections increase with decrease of the isotope mass number and with the energy of incident particles increase. On the other hand, the cross section ratios for tungsten isotopes to one of them become approximately constant with the incident nucleon energy increase. The values of these ratios were used for estimation of the difference in the fission barriers for neighbouring isotopes: $B_f(A-1)-B_f(A)$. Using the same approach which we applied in ref. [13] for lead isotopes, we have obtained the following values: ≈ 0.55 MeV for neutrons and ≈ 0.50 MeV for protons. These differences are close to ones obtained for the lead isotopes (0.38 MeV for neutrons and ≈ 0.30 MeV for protons) in ref. [13] and seem to be too large for liquid-drop fission. At present it is hard to make a conclusion whether this fact is a result of coarseness of the present suppositions (simplified relation for fission cross section, averaging on wide spectrum of fissioning nuclei) or it has a physical sense. More detailed calculations are necessary.

## 4 Comparison of the (p,f) and (n,f) cross sections

The dependence of the (p,f)/(n,f) cross-section ratio on the parameter $Z^2/A$ of a target nuclei was first studied in ref. [14] for a range of nuclei from Ta to Np. It has been observed that the $\sigma_{pf}/\sigma_{nf}$ ratio depends strongly on the incident nucleon energy in the low-energy region (20 – 70 MeV) and then approaches a plateau slowly. In fig. 2, up-to-date results for the incident nucleon energy of about 180 MeV are shown. It is seen that the $\sigma_{pf}/\sigma_{nf}$ ratio increases with the $Z^2/A$ parameter decrease. However, it is also seen that the dependency is not monotonous, but it has a "hollow" in transition from the lead isotopes group to the tungsten one. This effect has been predicted earlier in ref. [15] and was ascribed to connection of the fission cross-section ratio with a value of the fission barrier. The fission barrier changes weakly for heavy actinide nuclei, increases rapidly in the region of lighter nuclei and reaches a maximum value for $^{208}$Pb due to large value of the shell correction to the liquid-drop barrier. Since the shell correction decreases, the barrier reaches a plateau that can lead to slowing down of the $\sigma_{pf}/\sigma_{nf}$ rise in the region of tungsten.

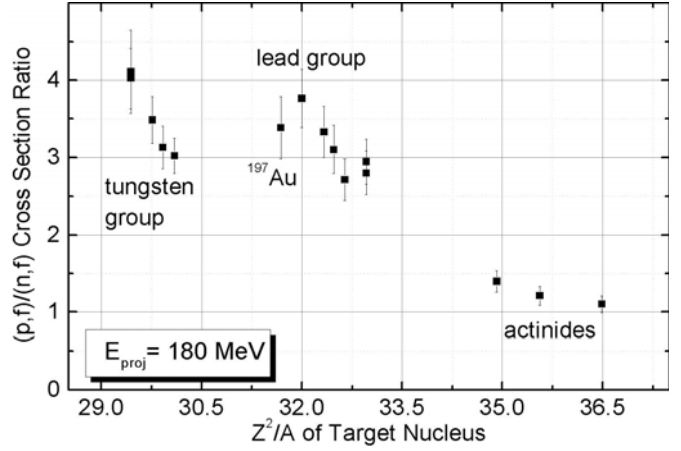

Fig. 2. The (p,f)/(n,f) cross-section ratios versus parameter $Z^2/A$ of the target nucleus.

## 5 Effect of "compound nucleus" in intermediate energy region.

Comparative analysis of data on (p,f) and (n,f) cross-sections for $^{205}$Tl, $^{204,206-208}$Pb and $^{209}$Bi in the energy range up to about 180 MeV [10] and results of the present work has given an additional confirmation of a regularity we recently observed in ref. [16], which can be interpreted as an effect of "compound nucleus" in intermediate energy region. The regularity is an equality of fission cross sections (within experimental errors, 10-20%,) in reactions of nuclei with protons and neutrons, characterizing by composite systems with the same sum charge, Z, and mass, A, at the same energy of incident particles.

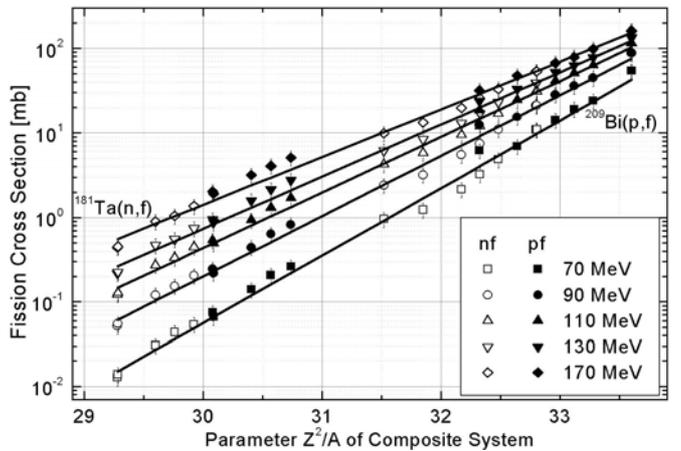

**Fig. 3.** The (n,f) and (p,f) cross-sections versus parameter $Z^2/A$ of the composite system for the different incident nucleon energies.

In fig. 3 the dependences are presented of the (n,f) and (p,f) cross-sections for nuclei from $^{181}$Ta to $^{209}$Bi on the parameter

$Z^2/A$ of a composite system. It is seen from fig. 3 an exponential character of the dependences - straight lines in semi-logarithmic scale, the slope of which is decreases with the projectile energy increase. Thus the dependency of the fission cross section values in these reactions is traced analogously to the dependence of the fission probability on the parameter $Z^2/A$ of the compound nucleus:

$$P_f \sim exp\{-[B_f(Z^2/A) - B_n]/T\},$$

where $B_f$ and $B_n$ are the fission barrier and neutron binding energy, respectively, and $T$ is the temperature of an excited nucleus. Increase of the temperature, which results from increasing projectile energy, thus leads to a weakening of the dependence of the fission probability on the parameter $Z^2/A$

The effect is seen in other earlier observed properties of the fission process induced by intermediate energy nucleons too. In particular, as it was already discussed in sect. 4, the (p,f) cross sections exceed the (n,f) ones systematically. Fission fragment angular anisotropy in intermediate energy region show a similar behaviour for proton- and neutron-induced fission [17].

Recently, measured fragment mass distributions in the fission of $^{232}$Th, $^{238}$U, $^{235}$U and $^{237}$Np, induced by protons with energies of 50 and 96 MeV, have been compared with calculations from the code TALYS [18, 19] and experimental data on neutron-induced fission of $^{238}$U for the neutron energy range up to 200 MeV [20]. It has been shown that the most striking characteristic of the fragment mass distribution – the ratio of symmetric and asymmetric fission fractions – in the case of neutrons is manifested in the same way as in the case of protons, i.e., at the same particle energy the fraction of symmetric fission drops with increasing number of neutrons in the composite nucleus, N: from N=142-143 (for $^{232}$Th+p, $^{235}$U+p, $^{237}$Np+p) to N=146 ($^{238}$U+p) and further to N=147 ($^{238}$U+n). Such a result was established earlier for protons and neutrons at an energy of about 15 MeV (leading to excitation energies above the barrier, near 14 MeV) [21], when almost each interaction of a proton or neutron with the target leads to a veritable compound nucleus, the decay mode of which does not depend on the way of formation.

The above mentioned experimental manifestations of the effect of "compound nucleus", pointing to a large resemblance of the behaviour of composite systems formed at intermediate energies and real compound nuclei at low energies, can be understood in the framework of the modern model codes. As a result of the first stage of the interaction of intermediate-energy nucleons with a nucleus (i.e., a cascade, followed by pre-equilibrium nucleon emission), a variety of nuclei is created, with different charges, masses and excitation energies, $W(Z_i, A_i, E_i^*)$. These nuclei reach thermal equilibrium, (i.e., compound nucleus formation), and finally undergo fission, thus contributing to the observed characteristics of fission. Calculations for the two reactions $^{208}$Pb +p and $^{209}$Bi+n performed in frame of the codes TALYS [16] and CEM03.01 [17] show that the average values $\langle Z \rangle$, $\langle A \rangle$, and, $\langle E^* \rangle$, versus the incident nucleon energy as well as the shapes of their distributions are very close to each other.

This work was performed in framework of the ISTC project 2213 and partially supported by U.S. DOE.